\documentclass[aps,twocolumn,showpacs,superscriptaddress,preprintnumbers]{revtex4}
\usepackage{amssymb}
\usepackage{amsmath}
\usepackage{graphicx}
\usepackage{epsfig}

\input{tcilatex}

\begin{document}

\title{Localization of Relative-Position of Two Atoms Induced by Spontaneous
Emission}
\author{L.~Zheng}
\affiliation{The Center of Theoretical Physics,Jilin University, Changchun,130023, China}
\affiliation{Institute of Theoretical Physics, The Chinese Academy of Science, Beijing,
100080, China}
\author{C.~Li}
\affiliation{Institute of Theoretical Physics, The Chinese Academy of Science, Beijing,
100080, China}
\author{Y.~Li}
\affiliation{Institute of Theoretical Physics, The Chinese Academy of Science, Beijing,
100080, China}
\author{C.~P.~Sun}
\email{suncp@itp.ac.cn; URL: http://www.itp.ac.cn/~suncp}
\affiliation{Institute of Theoretical Physics, The Chinese Academy of Science, Beijing,
100080, China}
\date{January 14, 2005}

\begin{abstract}
We revisit the back-action of emitted photons on the motion of the relative
position of two cold atoms. We show that photon recoil resulting from the
spontaneous emission can induce the localization of the relative position of
the two atoms through the entanglement between the spatial motion of
individual atoms and their emitted photons. The result provides a more
realistic model for the analysis of the environment-induced localization of
a macroscopic object.
\end{abstract}

\maketitle

\section{Introduction}

\noindent To understand the transition from the quantum world to the
classical world, one of the central issues is to consider how the
macroscopic object is localized in a certain spatial domain \cite{Joos,Omnes}%
. Superposition and its many particle version-the entanglement are the
essential features of quantum physics that permits macroscopic objects to
spread across the whole space, while classical physics is based on a local
realism and thus a macroscopic object in classical world has well-defined
position. The theory of quantum decoherence is a successful description
about the quantum-classical transition and it is explained that the loss of
quantum coherence of macroscopic objects is due to their coupling with the
environments. Therefore a perfect knowledge of the mechanism of decoherence
is crucial for the understanding of the quantum-classical transition since
the de-localization usually results from quantum coherence. On the other
hand in the science of quantum information, the information is mainly
processed by using the quantum coherence. Knowing how decoherence destroys
the wave nature of the matter in the wave-particle duality make it possible
to find decoherence-free states as the computation space \cite{zhan}.

Theoretical studies in this context have concerned a variety of modelled
systems \cite{zeh,zurek81,zur82,bose99,sun01,Zha02}. The corresponding
experiments have also been done in the last years\cite{brun,myatt,fridmen}
to demonstrate the dynamic process of decoherence, the collapse and revival
of quantum coherence. We studied the phenomenon of quantum decoherence of a
macroscopic object by introducing a novel concept, the adiabatic quantum
entanglement between collective states (such as that of the center-of-mass
(C.M)) and inner states \cite{sun01,sun-ad}. In the adiabatic separation of
slow and fast variables of a macroscopic object, its wave function can be
written as an entangled state with correlation between adiabatic inner
states and quasi-classical motional configuration of the C.M. Since the
adiabatic inner states are factorized with respect to the composing parts of
the macroscopic object \cite{sun93}, this adiabatic separation can induce
quantum decoherence for the collective motion. This observation thus
provides us with a possible solution to the Schr\"{o}dinger cat paradox at
least in the model level. Here, quantum-classical transition is just
characterized by the localization of macroscopic object. When this idea was
generalized to a triple system (the measured system together with the
pointer state of Schr\"{o}dinger-cat-like mater and its inner variables )
similar to that by Zurek \cite{zurek81,zur82}, a consistent approach for
quantum measurement was present by Zhang, Liu and Sun \cite{Zha02}.

Now a next-step is naturally to investigate the actual physical systems
(such as atoms interacting with the vacuum) illustrating the essence of such
environment-induced decoherence. Here, to focus on the essence of problem we
need not to consider the realistic macroscopic object consisting of too many
particles. In principle the localization phenomenon of the relative
coordinate of two atoms induced by environment is sufficient to account for
the fundamental conception behind such quantum decoherence problem.

Actually, people have studied a more realistic model involving a sequence of
external scattering interactions with a system of two particles considering
neither the inter-particle interaction \cite{Rau03} nor the inner structure
of particle. It shows that the scattering interactions progressively
entangles two particles and decohers their relative phase, naturally leading
to the localization of the particles in relative-position space. More
profound result for the measurement-induced localization has been discovered
and the phenomenon of phase entanglement is defined first in ref. \cite%
{Cha04}. It is found that there is phase entanglement only in the
coordinate-space, which can interpret spatial localization phenomenon of
atom. They also make a Schmidt-mode analysis of the entanglement between the
emitting atom and the emitted photon generated in the process of the
spontaneous emission and show that the localization of phonon can be
controlled by measuring the atom state\cite{Cha02}. Newly a model of two
entangled atoms located inside two spatially separated cavities has also
been investigated \cite{Yu04}. It is found that the local decoherence takes
an infinite time and the disentanglement due to spontaneous emission may
take a finite time. For the application in quantum computing, You has
investigated decoherence effects due to motional degrees of freedom of
trapped electronically coded atomic or ionic qubits\cite{You}.

In this article, we continue the studies of the environment-induced
decoherence for the motion of the C.M of a pair atoms induced by the
back-action of light emitted from the atomic inner states. Our investigation
will emphasize the reality of physical model examining the localization of
relative position due to such spontaneous emission. Under the second order
approximation we study in detail the time evolution of the C.M relevant
state by considering the realistic environment formed by the photons in
spontaneous emission, which causes the atomic recoils. The corresponding
localization phenomena is characterized by the spatial reduced density
matrix in the real space as the vanishing of the off-diagonal elements. This
paper is organized as follows. In section II, by neglecting multi-photon
processes as the higher order approximation we present a simplified model to
study the time evolution of the spatial states of the two atom system under
the back-action of emitted photons. In section III the spatial decoherence
induced by atomic spontaneous emission is studied by the caculation of the
reduced density matrix. Section IV demonstrates the localization of a
macroscopic object resulting from the spatial decoherence by two simple
examples, and finally conclusions are given in section V.

\section{Motion of C.M\ of \ two atoms influenced by vacuum electromagnetic
field}

\noindent Our system consists of a pair of noninteracting two-level atoms of
same mass $m$ and same transition frequency $\omega _{0}$ placed in the
vacuum electromagnetic field. Here and infra we use black body text to
denote vector quantities for convenient. The atoms are spatially separated
in the positions $\mathbf{r}_{A}$and $\mathbf{r}_{B}$ respectively and the
corresponding momentums are $\mathbf{p}_{A}$and $\mathbf{p}_{B}$ (as
illustrated in FIG.1). We denote the C.M and relative momentums respectively
by$\mathbf{P}=\mathbf{p}_{A}+\mathbf{p}_{B}$and $\mathbf{p}=(\mathbf{p}_{A}-%
\mathbf{p}_{B})/2,$ and similarly the C.M and relative positions can be
denoted respectively by $\mathbf{X}$ and $\mathbf{r}$.

Under the rotating wave approximation, the Hamiltonian of our system reads:

\begin{eqnarray}
H &=&\frac{\mathbf{P}^{2}}{2M}+\frac{\mathbf{p}^{2}}{2\mu }+\frac{1}{2}\hbar
\omega _{0}\left( \sigma _{z}^{(A)}+\sigma _{z}^{(B)}\right)  \notag \\
&&+\underset{\mathbf{k}}{\sum }\hbar \omega _{k}a_{\mathbf{k}}^{\dagger }a_{%
\mathbf{k}}+\hbar \underset{\mathbf{k}}{\sum }g(\mathbf{k})[(\sigma
_{+}^{(A)}e^{i\mathbf{k}\cdot (\mathbf{X}+\frac{\mathbf{r}}{2})}  \notag \\
&&+\sigma _{+}^{(B)}e^{i\mathbf{k}\cdot (\mathbf{X}-\frac{\mathbf{r}}{2}%
)})a_{\mathbf{k}}+h.c],  \label{Hamiton}
\end{eqnarray}%
where $M=2m$ and $\mu =m/2$. The atomic transition operator are denoted by $%
\sigma _{+}^{(i)}=\left\vert e_{i}\right\rangle \left\langle
g_{i}\right\vert $ and $\sigma _{-}^{(i)}=\left\vert g_{i}\right\rangle
\left\langle e_{i}\right\vert $ ($i=A,B$) with respect to the excited states
$\left\vert e\right\rangle $ and the ground states $\left\vert
g\right\rangle $ of each atom. $a_{\mathbf{k}}^{\dagger }$ and $a_{\mathbf{k}%
}$ are the annihilation and creation operators of the vacuum electromagnetic
fields mode $\mathbf{k}$ with frequency $\omega _{k}=ck$ ($k=\left\vert
\mathbf{k}\right\vert $). The coupling constant
\begin{equation}
\hbar g(\mathbf{k})=\sqrt{\frac{\hbar \omega _{k}}{2\varepsilon _{0}V}}%
\mathbf{\varepsilon }_{\mathbf{k}}\cdot \mathbf{d}
\end{equation}%
$\ $depends on the effective mode volume $V$, the polarization vector of the
field vector $\mathbf{\varepsilon }_{\mathbf{k}}$and the transition dipole
moment of the atom $\mathbf{d}$ .
\begin{figure}[tbp]
\includegraphics[bb=50 130 550 730, width=6cm,clip]{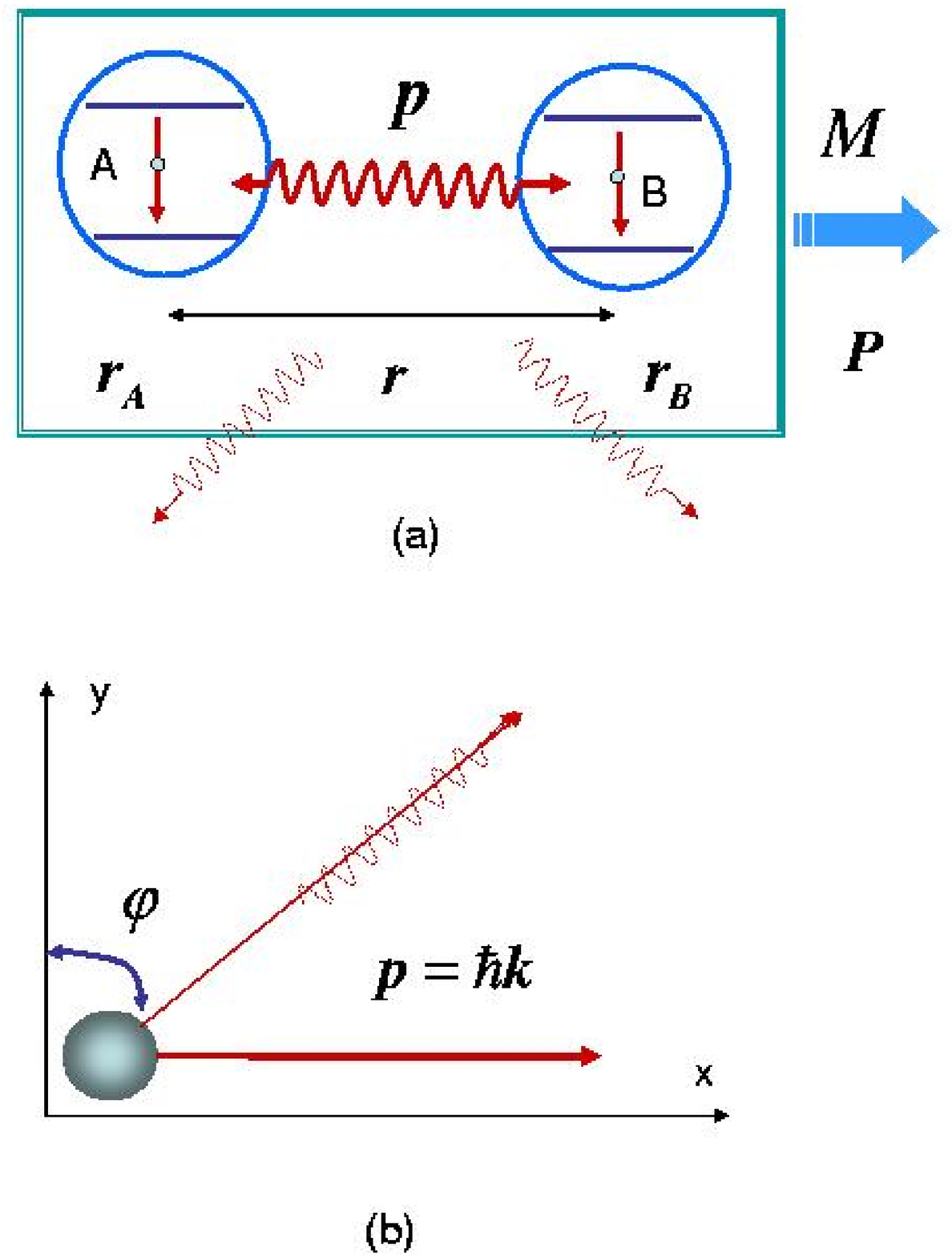}
\caption{\textit{(a). Schematic illustration of two atoms A and B in the
positions $\mathbf{r}_{A}$ and $\mathbf{r}_{B}$ interacting with vacuum
electromagnetic field coherently. (b). The vacuum light scattering in a
two-dimensional space: the atoms move only along the $x$ axis while the
photon can be emitted along the angle $\protect\varphi $ between the wave
vector $\mathbf{k}$ of the emitted photon and the $y$ axis. The momentum
kick given by the emitted photon to the atom is $p_{\protect\varphi }=\hbar
k\sin \protect\varphi .$} }
\end{figure}

For simplicity, we consider the problem in a two-dimensional space. We
assume the atoms move only along the $x$ axis while photons can be emitted
along any direction into the field. We also suppose $\varphi $ is the angle
between the wave vector $\mathbf{k}$ of the emitted photon and the $y$ axis.
The momentum kick given by the emitted photon to the atom is $p_{\varphi
}=\hbar k\sin \varphi .$ At $t=0$ the relative momentum of the atoms is
undefined and its state is a linear combination (superposition) of the
eigenstates $\left\vert p\right\rangle $ of relative momentum operator $%
\mathbf{p}$. The total momentum state is $\left\vert P\right\rangle $
corresponding to total momentum $\mathbf{P}$ and the two atoms are both in
excited states initially. The initial wave function of the system can be
written as a product state,

\begin{equation}
\left\vert \Psi (0)\right\rangle =\left\vert P\right\rangle \otimes \int
d^{3}pC_{p}\left\vert p\right\rangle \otimes \left\vert e_{1}\right\rangle
\otimes \left\vert e_{2}\right\rangle \otimes \left\vert 0\right\rangle
\label{initial state}
\end{equation}%
where $C_{p}$ is the distribution function corresponding to the relative
momentum eigenstate $\left\vert p\right\rangle $ and satisfies the
normalization condition $\int_{-\infty }^{\infty }\left\vert
C_{p}\right\vert ^{2}dp=1$, $\left\vert e_{1}\right\rangle $ and $\left\vert
e_{2}\right\rangle $ represent the excited states of the atom A and B
respectively. $\left\vert 0\right\rangle $ means the vacuum state of the
electromagnetic filed. The time evolution of the state $\left\vert \Psi
\left( t\right) \right\rangle $ is described by the Schr\"odinger equation.

To the second order approximation about the weak coupling characterized by $%
g(\mathbf{k})$, the state vector $\left\vert \Psi \left( t\right)
\right\rangle $ can be calculated as

\begin{eqnarray}
\left\vert \Psi \left( t\right) \right\rangle &=&\exp (-i\omega _{0}t)[\int
dpA\left( p,t\right) \left\vert P,p,e_{1},e_{2},0\right\rangle +  \notag
\label{wave function at t} \\
&&\underset{\mathbf{k}}{\sum }\int dpB_{\mathbf{k,}p}\left( t\right) \left[
\left\vert P-p_{\varphi },p-\frac{1}{2}p_{\varphi },g_{1},e_{2},1_{\mathbf{k}%
}\right\rangle \right.  \notag \\
&&+\left. \left\vert P-p_{\varphi },p+\frac{1}{2}p_{\varphi },e_{1},g_{2},1_{%
\mathbf{k}}\right\rangle \right]  \notag \\
&&+\underset{\mathbf{k,k}^{\prime }}{\sum }\int dpD_{\mathbf{k,k}^{\prime
},p}\left( t\right) \left\vert P-p_{\varphi }-p_{\varphi }^{\prime
}\right\rangle  \notag \\
&&\otimes \left\vert p-\frac{1}{2}(p_{\varphi }-p_{\varphi }^{\prime
})\right\rangle \otimes \left\vert g_{1},g_{2}\right\rangle \otimes
\left\vert 1_{\mathbf{k}}1_{\mathbf{k}^{\prime }}\right\rangle ]
\end{eqnarray}%
with $p_{\varphi }^{\prime }=\hbar k^{\prime }\sin \varphi ^{\prime }$ and $%
\int dp$\textbf{\ }roughly denotes the definite integral of $\int_{-\infty
}^{\infty }dp$. We notice that $\exp (-i\omega _{0}t)$ is a common phase
factor, and the first term in Eq. (\ref{wave function at t}) means that both
atoms are in the excited states while the field is in the state of vacuum.
The second term denotes one of the atoms decays to the ground state $%
\left\vert g_{i}\right\rangle $ ($i=1,2$) from the excited states $%
\left\vert e_{i}\right\rangle $ with a photon of momentum $\hbar \mathbf{k}$
emitted simultaneously. The last term describes the situation when both
atoms jump down to the ground states emitting two photons with momentum $%
\hbar \mathbf{k}$ and $\hbar \mathbf{k}^{\prime }$ respectively.

The time-dependent coefficients $A\left( p,t\right) $, $B_{\mathbf{k,}%
p}\left( t\right) $ and $D_{\mathbf{k,k}^{\prime },p}\left( t\right) $ can
be calculated by directly solving the Schr\"odinger equation. The obtained
system of equations are

\begin{equation}
\overset{\cdot }{A}_{p}+i\left( \omega _{A}-\omega _{0}\right) A_{p}=-2i%
\underset{\mathbf{k}}{\sum }B_{\mathbf{k,}p}\left( t\right) g(\mathbf{k}),
\label{differential eguation for A}
\end{equation}

\begin{eqnarray}
&&\overset{\cdot }{B}_{\mathbf{k,}p}+[i\omega _{B}\left( \mathbf{k}\right)
-\omega _{0}]B_{\mathbf{k,}p}=  \notag \\
&&-iA_{p}g(\mathbf{k})-i\underset{\mathbf{k}^{\prime }}{\sum }D_{\mathbf{k,k}%
^{\prime },p}g(\mathbf{k}^{\prime })  \label{differential eguation for B}
\end{eqnarray}%
and%
\begin{eqnarray}
&&\overset{\cdot }{D}_{\mathbf{k,k}^{\prime },p}+[i\omega _{D}(\mathbf{k,k}%
^{\prime })-\omega _{0}]D_{\mathbf{k,k}^{\prime },p}=  \notag \\
&&-iB_{\mathbf{k,}p}g(\mathbf{k}^{\prime })-iB_{\mathbf{k}^{\prime }\mathbf{,%
}p}g(\mathbf{k}),  \label{differential eguation for C}
\end{eqnarray}%
where the coefficients
\begin{eqnarray}
\hbar \omega _{A} &=&P^{2}/2M+p^{2}/2\mu +\hbar \omega _{0},  \label{wa} \\
\hbar \omega _{B}\left( \mathbf{k}\right) &=&\frac{\left( P-p_{\varphi
}\right) ^{2}}{2M}+\frac{(p-\frac{1}{2}p_{\varphi })^{2}}{2\mu }+\hbar
\omega _{k}  \label{wb2}
\end{eqnarray}%
describe the scattering processes with photon recoil while
\begin{eqnarray}
\hbar \omega _{D}\left( \mathbf{k,k}^{\prime }\right) &=&\frac{\left(
P-p_{\varphi }-p_{\varphi }^{\prime }\right) ^{2}}{2M}  \notag \\
&&+\frac{\left( p-\frac{1}{2}p_{\varphi }+\frac{1}{2}p_{\varphi }^{\prime
}\right) ^{2}}{2\mu }  \notag \\
&&+\hbar \omega _{k}+\hbar \omega _{k^{\prime }}-\hbar \omega _{0}
\label{wd}
\end{eqnarray}%
means that the two photon scattering will induce the momentum transfer.
Notice that in the above calculation we have ignored the higher order
multi-photon processes.

Starting with the initial conditions $A\left( p,0\right) =C_{p}$, $B_{%
\mathbf{k,}p}\left( 0\right) =0$ and $D_{\mathbf{k,k}^{\prime },p}\left(
0\right) =0$, the Laplace transformation about the above system of equations
can give the explicit solutions to $A_{p}\left( p,t\right) $, $B_{\mathbf{k,}%
p}\left( t\right) $ and $D_{\mathbf{k,k}^{\prime },p}\left( t\right) $ in
the Weisskopf-Wigner approximation \cite{louisell} (see Appendix for
detailed calculations). For the purpose of this paper we need not write down
them here for arbitrary time $t.$ In the limit $t\rightarrow \infty $, we
have $A_{p}\left( \infty \right) \rightarrow 0$, $B_{\mathbf{k,}p}\left(
\infty \right) \rightarrow 0$ and%
\begin{eqnarray}
&&D_{\mathbf{k,k}^{\prime },p}\left( \infty \right) \rightarrow \frac{g(%
\mathbf{k}^{\prime })g(\mathbf{k})C_{p}}{\left( i\left( \omega _{B}\left(
\mathbf{k}\right) -\omega _{D}\left( \mathbf{k,k}^{\prime }\right) \right) +%
\frac{\Gamma }{2}\right) }  \notag \\
&&\times \frac{e^{-i\left( \omega _{D}\left( \mathbf{k,k}^{\prime }\right)
-\omega _{0}\right) t}}{\left( i\left( \omega _{B}\left( \mathbf{k}^{\prime
}\right) -\omega _{D}\left( \mathbf{k,k}^{\prime }\right) \right) +\frac{%
\Gamma }{2}\right) }
\end{eqnarray}%
where $\Gamma $ is the decay rate of an atom from state $\left\vert
e\right\rangle $ to state $\left\vert g\right\rangle $. Neglecting the small
recoil energies, we have

\begin{eqnarray}
&&D_{\mathbf{k,k}^{\prime },p}\left( \infty \right) \rightarrow \frac{g(%
\mathbf{k}^{\prime })g(\mathbf{k})C_{p}}{i\left( \omega _{0}-\omega
_{k^{\prime }}-\left( \frac{p}{2\mu }-\frac{P}{M}\right) \frac{p_{\varphi
}^{\prime }}{\hbar }\right) +\frac{\Gamma }{2}}  \notag \\
&&\times \frac{e^{-i\left( \omega _{D}\left( \mathbf{k,k}^{\prime }\right)
-\omega _{0}\right) t}}{i\left( \omega _{0}-\omega _{k}-\left( \frac{p}{2\mu
}-\frac{P}{M}\right) \frac{p_{\varphi }}{\hbar }\right) +\frac{\Gamma }{2}}
\label{ddsi}
\end{eqnarray}

In general the C.M momentum of a hot atom is very large and so its momentum
exchange with the electromagnetic field can be neglected. In this sense the
electromagnetic field does not influence its C.M state nearly. However, it
is not the case for the ultracold atoms because their C.M momentums are very
small. The influence of the interaction between the atoms and the
electromagnetic field on the spatial motion of the atoms becomes very
important. Therefore it is an crucial issue about how the spatial states of
the ultracold atoms are affected by their electromagnetic field
environments. In the following section, we will go on to study how the
spontaneous emission affects the distribution of the atomic relative
position.

\section{The spatial decoherence induced by incoherent spontaneous Emission}

According to the above analysis, after a sufficiently long time $t>>1/\Gamma
$, the state of the system becomes%
\begin{eqnarray}
\left\vert \Psi \right\rangle &\rightarrow &\underset{\mathbf{k,k}^{\prime }}%
{\sum }\int dpD_{\mathbf{k,k}^{\prime },p}\left( \infty \right) \left\vert
P-p_{\varphi }-p_{\varphi }^{\prime }\right\rangle \otimes  \notag \\
&&\left\vert p-\frac{1}{2}(p_{\varphi }-p_{\varphi }^{\prime })\right\rangle
\otimes \left\vert g_{1},g_{2}\right\rangle \otimes \left\vert 1_{\mathbf{k}%
}1_{\mathbf{k}^{\prime }}\right\rangle .  \label{psi}
\end{eqnarray}%
Supposing that the modes of field are closely spaced in frequency domain, we
can replace $\underset{\mathbf{k,k}^{\prime }}{\sum }$ by the integral of
\begin{equation*}
\frac{V^{2}}{\left( 2\pi \right) ^{4}}\int_{0}^{\infty }kdk\int_{0}^{2\pi
}d\varphi \int_{0}^{\infty }k^{\prime }dk^{\prime }\int_{0}^{2\pi }d\varphi
^{\prime }.
\end{equation*}%
Considering that the velocity of a realistic atom is far smaller than the
light velocity in vacuum, we can further simplify Eq. (\ref{psi}) in the
representation of the C.M- relative coordinates ($X$ and $x)$:%
\begin{eqnarray}
\left\vert \Psi \right\rangle &=&N\int d\xi e^{\frac{i}{\hbar }px}e^{-\frac{i%
}{2\hbar }x(p_{\varphi }-p_{\varphi }^{\prime })}\left\vert
X,x,g_{1},g_{2},1_{k,\varphi }1_{k^{\prime },\varphi ^{\prime }}\right\rangle
\notag \\
&&\times \frac{C_{p}e^{\frac{i}{\hbar }PX}e^{-\frac{i}{\hbar }X(p_{\varphi
}^{\prime }+p_{\varphi })}e^{-i\left( \omega _{D}\left( \mathbf{k,k}^{\prime
}\right) -\omega _{0}\right) t}}{\left[ i\left( \omega _{0}-\omega
_{k^{\prime }}\right) +\frac{\Gamma }{2}\right] \left[ i\left( \omega
_{0}-\omega _{k}\right) +\frac{\Gamma }{2}\right] }
\end{eqnarray}%
where we use $\int d\xi $\textbf{\ }roughly denotes the definite
multi-integral of\textbf{\ }%
\begin{equation*}
\int_{0}^{\infty }dk\int_{0}^{2\pi }d\varphi \int_{0}^{\infty }dk^{\prime
}\int_{0}^{2\pi }d\varphi ^{\prime }\int_{-\infty }^{\infty }dx\int_{-\infty
}^{\infty }dX\int_{-\infty }^{\infty }dp
\end{equation*}%
\textbf{\ }and $N$ is a normalization factor including the slowly varying $g(%
\mathbf{k}^{\prime })$ and $g(\mathbf{k})$.

Tracing over the variables of the electromagnetic field, the inner states of
the atoms and the C.M motion, one can obtain the reduced density matrix

\begin{eqnarray}
&&\rho \left( x,x^{\prime },t\right) =  \notag \\
&&N^{\prime }\int d\varphi d\varphi ^{\prime }e^{-i\frac{\omega _{0}}{2c}%
\left( x-x^{\prime }\right) \sin \varphi }e^{i\frac{\omega _{0}}{2c}\left(
x-x^{\prime }\right) \sin \varphi ^{\prime }}  \notag \\
&&\times \psi \left( x+\frac{\hbar \omega _{0}t\left( \sin \varphi +\sin
\varphi ^{\prime }\right) }{2\mu c},t\right)  \notag \\
&&\times \psi ^{\ast }\left( x^{\prime }+\frac{\hbar \omega _{0}t\left( \sin
\varphi +\sin \varphi ^{\prime }\right) }{2\mu c},t\right)  \notag \\
&\approx &N^{\prime }\psi \left( x,t\right) \psi ^{\ast }\left( x^{\prime
},t\right) J_{0}\left[ \frac{\omega _{0}}{2c}\left( x-x^{\prime }\right) %
\right] ^{2}  \label{DEN}
\end{eqnarray}%
where $N^{\prime }$ is the normalization factor and
\begin{equation}
\psi \left( x,t\right) =\int_{-\infty }^{\infty }C_{p}\exp (\frac{i}{\hbar }%
\left( px-\frac{p^{2}}{2\mu }t\right) )dp.  \label{wav fun}
\end{equation}%
In Eq.(\ref{DEN}), considering that the term of $\hbar \omega _{0}t\left(
\sin \varphi +\sin \varphi ^{\prime }\right) /\left( 2\mu c\right) $ means
the small offset induced by atomic spontaneous emission of the relative
position between the two atoms, we have expanded $\psi \left( x+\hbar \omega
_{0}t\left( \sin \varphi +\sin \varphi ^{\prime }\right) /\left( 2\mu
c\right) ,t\right) $ around $x$ to the first order approximation and also
supposed $\omega _{0}>>\Gamma $. $J_{0}\left( z\right) $ is the Bessel
function of the first kind. Here we have used the following integral formula
($a$ is real number):%
\begin{equation*}
\int_{0}^{2\pi }\cos (a\sin (z))dz=2\pi J_{0}\left( a\right) ,
\end{equation*}

Now we can define the decoherence factor $F(x,x^{\prime })$ as

\begin{equation}
F(x,x^{\prime })=J_{0}^{2}(\frac{\pi }{\lambda }(x-x^{\prime })).
\end{equation}%
Here $\lambda =2\pi c/\omega _{0}$ is the wave length of the atomic
radiation. The elements of the reduced density matrix can be rewritten as
\begin{equation}
\rho \left( x,x^{\prime },t\right) =N^{\prime }\psi \left( x,t\right) \psi
^{\ast }\left( x^{\prime },t\right) F(x,x^{\prime }).  \label{dsm}
\end{equation}%
\begin{figure}[tbp]
\includegraphics[bb=45 42 568 335, width=8cm,clip]{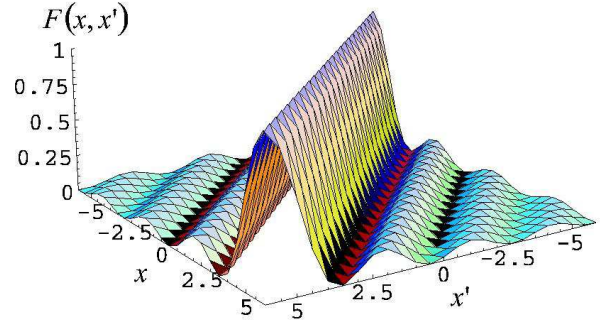}
\caption{Schematics of the decoherence factor $F(x,x^{\prime })$. When $%
\frac{\protect\pi }{\protect\lambda }\left( x-x^{\prime }\right) <e^{-1}$,
there exists a perfect quantum interference.When $F(x,x^{\prime })>e^{-3}$,
the quantum coherence disappears.}
\end{figure}

In FIG. 2, we draw \ the schematic curve of $F(x,x^{\prime })$. \ It is
illustrated that the off-diagonal elements of the reduced density matrix
decline to zero and thus the quantum coherence of system is lost. And the
diagonal elements of the reduced density matrix are suppressed with the
ultimate breadth $\lambda /\left( \pi e\right) $. The result also shows that
the quantum interference becomes more clear as the wave length of photon
emitted become larger or the distance of the atoms becomes smaller.

\section{From Decoherence to Localization of Macroscopic Object}

Now we take a simple example to illustrate how the above discussed
decoherence can result in the localization of a macroscopic object. We take
the initial state as a superposition

\begin{equation}
\Psi (0)=\psi \left( x\right) =\frac{1}{\sqrt{2}}\left(
G_{-}(x)+G_{+}(x)\right)  \label{wb}
\end{equation}%
of two Gaussian wave packets
\begin{equation}
G_{\pm }(x)=\frac{1}{\sqrt[4]{2\pi d^{2}}}\exp [-\frac{\left( x\pm a\right)
^{2}}{4d^{2}}].
\end{equation}%
As pointed out in ref.[1], the models with this initial state may arise in
the double slit experiment. Now we study the dynamical evolution of the wave
packets when $t>>1/\Gamma $. In the coordinate picture the elements of the
corresponding reduced density matrix can be expressed as
\begin{eqnarray}
\rho \left( x,x\prime ,t\right) &=&N^{\prime }\psi \left( x,t\right) \psi
^{\ast }\left( x^{\prime },t\right) F(x,x^{\prime })=  \notag \\
&&N^{\prime }J_{0}(\frac{\pi }{\lambda }(x-x^{\prime }))^{2}\left\{
G_{-}(x,t)G_{-}(x^{\prime },t)\right.  \notag \\
&&+G_{-}(x,t)G_{+}(x^{\prime },t)+G_{+}(x,t)G_{-}(x^{\prime },t)  \notag \\
&&\left. +G_{+}(x,t)G_{+}(x^{\prime },t)\right\} ,
\end{eqnarray}%
\ \ where%
\begin{eqnarray}
G_{+}(x,t) &=&\frac{1}{\left( 2\pi \right) ^{1/4}\sqrt{d+it\hbar /\left(
2\mu d\right) }}  \notag \\
&&\times \exp \left( -\frac{\left( 1-\frac{i\hbar }{2\mu d^{2}}t\right)
\left( x+a\right) ^{2}}{4d^{2}+\frac{t^{2}\hbar ^{2}}{\mu ^{2}d^{2}}}\right)
.  \label{g+0}
\end{eqnarray}

In fact, the spreading speed of the wave packets may be faster than the
speed of the decay of the atoms when the distance of the two atoms is far
smaller because the spreading speed is fast in the case. The wave packets
have overlapped before the atoms decay. We will discuss the case in the
following and here we suppose the distance of the atoms is large enough so
that we can say the two packets have not overlapped when we study it at some
time $t>>\frac{1}{\Gamma }$. We draw three groups of two-dimensional
schematics for the reduced density matrix at different times in FIG.3. The
first figure in each group is corresponding to the case when there is no
spontaneous emission and the other figure is corresponding to the case when
the influence of the spontaneous emission is considered. It demonstrates the
evolution of the localization of the two initial Gaussian wave packets.
\begin{figure}[tbp]
\includegraphics[height=12cm,width=9cm]{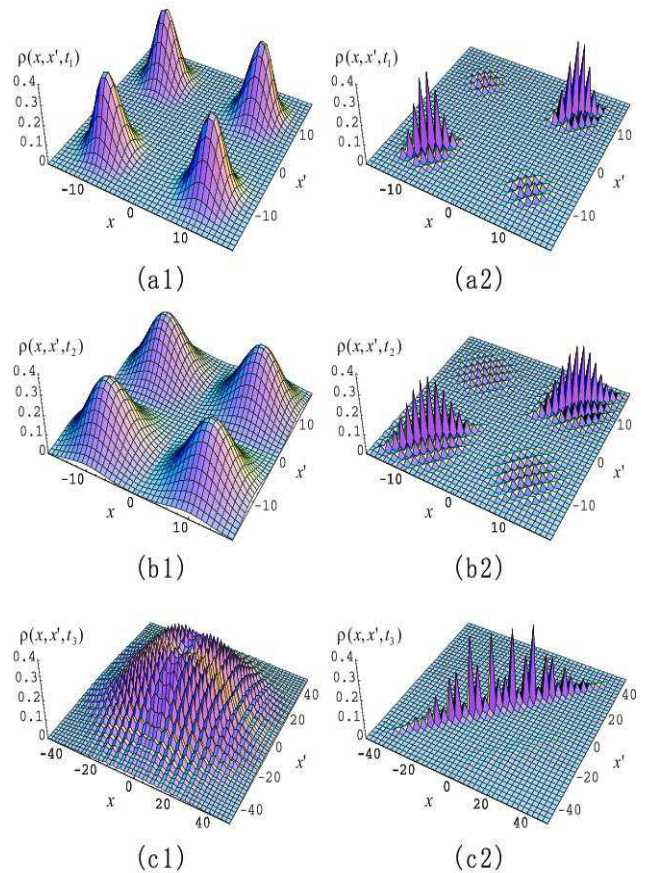}
\caption{Schematics of the evolution of the system state at three
different
times $t_{1}=\frac{100}{\Gamma },t_{2}=\frac{200}{\Gamma }$ and $t_{3}=\frac{%
1000}{\Gamma }$. The left three figures demonstrate the state evolution when
there is no spontaneous emission while the right three figures represent the
case when spontaneous emission exist. The third figure corresponds to the
case when the two packets have overlapped. These Schematic illustrates that
the quantum coherence of the system is lost in presente of spontaneous
emission and thus the spreading wave packets are suppressed.}
\label{fig.1}
\end{figure}

Actually, about ten years ago\cite{sun-yu}, we have generally studied a
quite simple dynamical problem: the motion of the wave packet for a `free'
particle of one dimension in present a dissipative environment. An
interesting result is that the dissipation suppresses the spreading of the
wave packet if the breadth of initial wave packet is so wide that the effect
of Brownian motion can be ignored. However, for the case with dissipation,
there appears to be a significant difference about the wave packet
spreading. This suppression of the wave packet spreading by dissipation
possibly provides a mechanism to localize the macroscopic object. It might
be of interest to note that the finite value of the width of the damped
particle wave packet for $t\rightarrow \infty $ leads to exactly the same
final value for the uncertainty product of the damped free particle, also
found by Schuch et al. using a nonlinear Schr\"{o}dinger equation\cite{shuch}%
. In the following, we will demonstrate such localization in our present
realistic model.

We take the initial state as a narrow Gaussian wave packet $G_{+}(x-a)$ for
the relative position representation of the two-atom system. Obviously the
narrowness of wave packet implies that the two atom system is initially in
localization. If there would not exist the spontaneous emission, the
relative position Gaussian wave packet would spread into the full space
infinitely and the localization of wave packet is lost during the evolution
of the system. Its breadth increases to infinity while its height decreases
from its initial value to zero. In present of the spontaneous emission, we
calculate the time evolution of
\begin{eqnarray}
G_{+}(x-a,t &=&0)=\int_{-\infty }^{\infty }C_{p}e^{i\frac{p}{\hbar }x}dp, \\
C_{p} &=&\frac{2d^{2}}{\pi \sqrt{2\pi }}e^{-\frac{d^{2}}{\hbar }p^{2}}.
\end{eqnarray}%
According to Eq. (\ref{wav fun}) and Eq. (\ref{dsm}), the reduced density
matrix at time $t>>\frac{1}{\Gamma }$ is

\begin{equation}
\rho \left( x,x^{\prime },t\right) =N^{\prime }G_{+}(x-a,t)G_{+}^{\ast
}(x^{\prime }-a,t)F(x,x^{\prime })  \label{rog+}
\end{equation}%
where%
\begin{eqnarray}
G_{+}(x-a,t) &=&\frac{1}{\left( 2\pi \right) ^{1/4}\sqrt{d+it\hbar /\left(
2\mu d\right) }}  \notag \\
&&\times \exp \left( -\frac{\left( 1-\frac{i\hbar }{2\mu d^{2}}t\right) x^{2}%
}{4d^{2}+\frac{t^{2}\hbar ^{2}}{\mu ^{2}d^{2}}}\right) .
\end{eqnarray}%
According to Eq. (\ref{rog+}), we can conclude that the breadth of the
spreading wave packet is suppressed as can also be seen in FIG. 4. We give
the schematics of the evolution of the wave packet at three different times
and in two cases: one is when there is no spontaneous emission and the other
is when there exists spontaneous emission. From the Eq. (\ref{rog+}) and
FIG. 4, we can see the wave packet spreading is suppressed and the ultimate
breadth is related to the wave length of the atomic radiation. Longer is the
atomic radiative wave length, wider is the breadth of the ultimate wave
packet.
\begin{figure}[tbp]
\includegraphics[height=12cm,width=9cm]{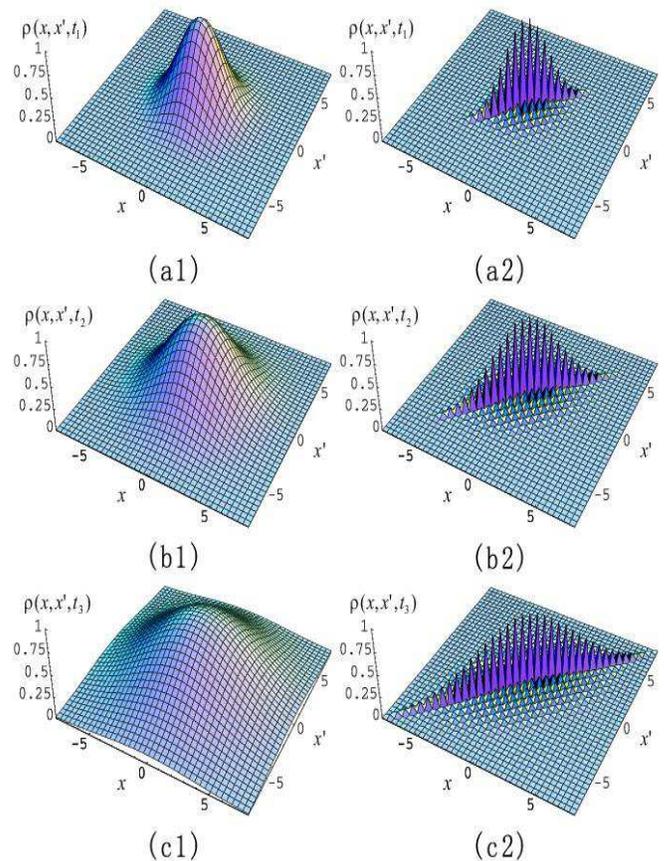}
\caption{Schematics of the suppression of a wave packet spreading.
The left three figures correspond to the evolution of the single
wave packet at three different instance when there is no
spontaneous emission and the right three
figures represent the cases when the spontaneous emissions exist. Here $%
t_{1}=\frac{2}{\Gamma },t_{2}=\frac{3}{\Gamma }$ and $t_{3}=\frac{5}{\Gamma }
$. They demonstrate how the atomic spontaneous emission suppresses the
spreading of wave packet.}
\end{figure}

\section{Conclusions}

\noindent In conclusion, we have investigated the atomic spontaneous
emission induced quantum decoherence phenomenon in association with the
localization of the relative position of a two atom system. The spontaneous
emission or the interaction with vacuum electromagnetic field may be a
fundamental process destroying the quantum effects in macroscopic objects.
By analyzing two simple examples, we demonstrate how the spontaneous
emission suppresses the spreading wave packet and thus localizes a
macroscopic object.

\textit{This work is supported by NSFC with Numbers 90203018, 10474104 and
10447133. It is also funded by the National Fundamental Research Program of
China with No. 2001CB309310. Z.L thanks for the helpful discussions with
Zhang Peng.}

\appendix*

\section{Solutions of the Schr\"{o}dinger equation}

In this appendix we give the calculations on solving the system of equations
consisting of Eq. (\ref{differential eguation for A}) to Eq. (\ref%
{differential eguation for C}). We take the Laplace transformation to Eq. (%
\ref{differential eguation for A}) to Eq. (\ref{differential eguation for C}%
) and obtain

\begin{eqnarray}
&&\left( L+i\left( \omega _{A}-\omega _{0}\right) \right) A\left( p,L\right)
\notag \\
&=&-2i\underset{\mathbf{k}}{\sum }B_{\mathbf{k,}p}\left( L\right) g(\mathbf{k%
})+C_{p},  \label{lt1}
\end{eqnarray}

\begin{eqnarray}
&&\left( L+i\left( \omega _{B}\left( \mathbf{k}\right) -\omega _{0}\right)
\right) B_{\mathbf{k,}p}\left( L\right)  \notag \\
&=&-iA\left( p,L\right) g(\mathbf{k})-i\underset{\mathbf{k}}{\sum }D_{%
\mathbf{k,k}^{\prime },p}\left( L\right) g(\mathbf{k}^{\prime }),
\label{lt2}
\end{eqnarray}

\begin{eqnarray}
&&\left[ L+i\left( \omega _{D}\left( \mathbf{k,k}^{\prime }\right) -\omega
_{0}\right) \right] D_{\mathbf{k,k}^{\prime },p}\left( L\right)  \notag \\
&=&-iB_{\mathbf{k,}p}\left( L\right) g(\mathbf{k}^{\prime })-iB_{\mathbf{k,}%
p}\left( L\right) g(\mathbf{k}).  \label{lt3}
\end{eqnarray}

From Eq. (\ref{lt3}), we have

\begin{equation}
D_{\mathbf{k,k}^{\prime },p}\left( L\right) =\frac{-iB_{\mathbf{k,}p}\left(
L\right) g(\mathbf{k}^{\prime })-iB_{\mathbf{k,}p}\left( L\right) g(\mathbf{k%
})}{L+i\left( \omega _{D}\left( \mathbf{k,k}^{\prime }\right) -\omega
_{0}\right) }.  \label{d}
\end{equation}%
Submitting Eq. (\ref{d}) into Eq. (\ref{lt2}), we can rewrite Eq. (\ref{lt2}%
) as

\begin{eqnarray}
&&\left( L+i\left( \omega _{B}\left( \mathbf{k}\right) -\omega _{0}\right)
\right) B_{\mathbf{k,}p}\left( L\right)  \notag \\
&=&-iA\left( p,L\right) g(\mathbf{k})  \notag \\
&&-B_{\mathbf{k,}p}\left( L\right) \sum_{\mathbf{k}^{\prime }}\frac{g^{2}(%
\mathbf{k}^{\prime })}{L+i\left( \omega _{D}\left( \mathbf{k,k}^{\prime
}\right) -\omega _{0}\right) }  \notag \\
&&-\sum_{\mathbf{k}^{\prime }}\frac{B_{\mathbf{k}^{\prime }\mathbf{,}%
p}\left( L\right) g(\mathbf{k}^{\prime })g(\mathbf{k})}{L+i\left( \omega
_{D}\left( \mathbf{k,k}^{\prime }\right) -\omega _{0}\right) }.  \label{b}
\end{eqnarray}%
Since we have ignored the higher order multi-photon processes, we can omit
the last term in the right of Eq. (\ref{b}). Then we obtain

\begin{equation}
B_{\mathbf{k,}p}\left( L\right) =\frac{-iA\left( p,L\right) g(\mathbf{k})}{%
L+i\left( \omega _{B}\left( \mathbf{k}\right) -\omega _{0}\right) +\underset{%
\mathbf{k}^{\prime }}{\sum }\frac{g^{2}(\mathbf{k}^{\prime })}{L+i\left(
\omega _{D}\left( \mathbf{k,k}^{\prime }\right) -\omega _{0}\right) }}.
\label{b2}
\end{equation}%
According to the Weisskopf-Wigner approximation\cite{louisell}, we can obtain

\begin{equation*}
\frac{\Gamma }{2}+i\triangle \omega =\underset{\mathbf{k}^{\prime }}{\sum }%
\frac{g^{2}(\mathbf{k}^{\prime })}{L+i\left( \omega _{D}\left( \mathbf{k,k}%
^{\prime }\right) -\omega _{0}\right) }
\end{equation*}%
where $\Gamma =\omega _{0}^{2}\left\vert \mathbf{d}\right\vert ^{2}/\left(
4\varepsilon _{0}\hbar c^{2}\right) $ is the decay rate of an atom from
state $\left\vert e\right\rangle $ to state $\left\vert g\right\rangle $ and
$\triangle \omega $ is the Lamb shift which is omitted in our following
calculations since it can be merged into the transition frequency $\omega
_{0}$. Eq. (\ref{b2}) can be simplified as

\begin{equation}
B_{\mathbf{k,}p}\left( L\right) =\frac{-iA\left( p,L\right) g(\mathbf{k})}{%
L+i\left( \omega _{B}\left( \mathbf{k}\right) -\omega _{0}\right) +\frac{%
\Gamma }{2}}.  \label{b3}
\end{equation}

Submitting Eq. (\ref{b3}) into Eq. (\ref{lt1}), we have

\begin{equation*}
A\left( p,L\right) =\frac{C_{p}}{L+i\left( \omega _{A}-\omega _{0}\right) +2%
\underset{\mathbf{k}}{\sum }\frac{g^{2}(\mathbf{k})}{L+i\left( \omega
_{B}\left( \mathbf{k}\right) -\omega _{0}\right) +\frac{\Gamma }{2}}}.
\end{equation*}%
\begin{equation}  \label{a2}
\end{equation}%
In the Weisskopf-Wigner approximation, we can also obtain%
\begin{equation*}
\frac{\Gamma }{2}+i\triangle \omega =\underset{\mathbf{k}}{\sum }\frac{g^{2}(%
\mathbf{k})}{L+i\left( \omega _{B}\left( \mathbf{k}\right) -\omega
_{0}\right) +\frac{\Gamma }{2}}.
\end{equation*}%
So Eq. (\ref{a2}) can be written as
\begin{equation}
A\left( p,L\right) =\frac{C_{p}}{L+i\left( \omega _{A}-\omega _{0}\right)
+\Gamma }.  \label{a3}
\end{equation}%
Combining Eq. (\ref{d}), Eq. (\ref{b3}) and Eq. (\ref{a3}), we can obtain
\begin{eqnarray}
B_{\mathbf{k,}p}\left( L\right) &=&\frac{-iC_{p}g(\mathbf{k})}{L+i\left(
\omega _{B}\left( \mathbf{k}\right) -\omega _{0}\right) +\frac{\Gamma }{2}}
\notag \\
&&\times \frac{1}{L+i\left( \omega _{A}-\omega _{0}\right) +\Gamma },
\label{b4}
\end{eqnarray}%
\begin{eqnarray}
D_{\mathbf{k,k}^{\prime },p}\left( L\right) &=&\frac{-g(\mathbf{k}^{\prime
})g(\mathbf{k})C_{p}}{L+i\left( \omega _{D}\left( \mathbf{k,k}^{\prime
}\right) -\omega _{0}\right) }  \notag \\
&&\times \frac{1}{L+i\left( \omega _{A}-\omega _{0}\right) +\Gamma }  \notag
\\
&&\times \left( \frac{1}{\left( L+i\left( \omega _{B}\left( \mathbf{k}%
\right) -\omega _{0}\right) +\frac{\Gamma }{2}\right) }\right.  \notag \\
&&+\left. \frac{1}{\left( L+i\left( \omega _{B}\left( \mathbf{k}^{\prime
}\right) -\omega _{0}\right) +\frac{\Gamma }{2}\right) }\right) .  \label{d2}
\end{eqnarray}%
Taking the inverse Laplace transformation to the above three equations, we
obtain the solutions to equations from Eq. (\ref{differential eguation for A}%
) to Eq. (\ref{differential eguation for C}):%
\begin{equation}
A\left( p,t\right) =C_{p}e^{-\Gamma t}e^{-i\left( \omega _{A}-\omega
_{0}\right) t},  \label{r1}
\end{equation}

\begin{equation}
B_{\mathbf{k,}p}\left( t\right) =-ig(\mathbf{k})C_{p}\frac{e^{(i\omega _{0}-%
\frac{\Gamma }{2}t)}\left( e^{-i\omega _{B}t}-e^{-i\omega _{A}t}\right) }{%
i\left( \omega _{A}-\omega _{B}\left( \mathbf{k}\right) \right) +\frac{%
\Gamma }{2}},
\end{equation}

\begin{eqnarray}
&&D_{\mathbf{k,k}^{\prime },p}\left( t\right)  \notag \\
&=&\frac{g(\mathbf{k}^{\prime })g(\mathbf{k})C_{p}}{i\left( \omega
_{A}-\omega _{D}\left( \mathbf{k,k}^{\prime }\right) \right) +\Gamma }
\notag \\
&&\times \lbrack \frac{e^{-i\left( \omega _{D}\left( \mathbf{k,k}^{\prime
}\right) -\omega _{0}\right) t}}{i\left( \omega _{B}\left( \mathbf{k}\right)
-\omega _{D}\left( \mathbf{k,k}^{\prime }\right) \right) +\frac{\Gamma }{2}}-%
\frac{e^{-i\left( \omega _{A}-\omega _{0}\right) t-\Gamma t}}{i\left( \omega
_{B}\left( \mathbf{k}\right) -\omega _{A}\right) -\frac{\Gamma }{2}}  \notag
\\
&&+\frac{e^{-i\left( \omega _{B}\left( \mathbf{k}\right) -\omega _{0}\right)
t-\frac{\Gamma }{2}t}}{\left[ i\left( \omega _{D}\left( \mathbf{k,k}^{\prime
}\right) -\omega _{B}\left( \mathbf{k}\right) \right) -\frac{\Gamma }{2}%
\right] \left[ i\left( \omega _{A}-\omega _{B}\left( \mathbf{k}\right)
\right) +\frac{\Gamma }{2}\right] }  \notag \\
&&+\frac{e^{-i\left( \omega _{D}\left( \mathbf{k,k}^{\prime }\right) -\omega
_{0}\right) t}}{i\left( \omega _{B}\left( \mathbf{k}^{\prime }\right)
-\omega _{D}\left( \mathbf{k,k}^{\prime }\right) \right) +\frac{\Gamma }{2}}-%
\frac{e^{-i\left( \omega _{A}-\omega _{0}\right) t-\Gamma t}}{i\left( \omega
_{B}\left( \mathbf{k}^{\prime }\right) -\omega _{A}\right) -\frac{\Gamma }{2}%
}  \notag \\
&&+\frac{e^{-i\left( \omega _{B}\left( \mathbf{k}\right) -\omega _{0}\right)
t-\frac{\Gamma }{2}t}}{\left[ i\left( \omega _{D}\left( \mathbf{k,k}^{\prime
}\right) -\omega _{B}\left( \mathbf{k}^{\prime }\right) \right) -\frac{%
\Gamma }{2}\right] \left[ i\left( \omega _{A}-\omega _{B}\left( \mathbf{k}%
^{\prime }\right) \right) +\frac{\Gamma }{2}\right] }].  \notag \\
\end{eqnarray}

\end{document}